\documentclass[aps, superscriptaddress, showpacs,prl, twocolumn]{revtex4-1}
\usepackage{ORI_Group_style}
\usepackage{blindtext}
\usepackage[dvipsnames]{xcolor}
\usepackage[normalem]{ulem}

\hypersetup{colorlinks,linkcolor=blue,citecolor=red,urlcolor=blue}

\begin{document}

\title{Quantum Acoustomechanics with a Micromagnet}

\author{Carlos Gonzalez-Ballestero}
\affiliation{Institute for Quantum Optics and Quantum Information of the
Austrian Academy of Sciences, A-6020 Innsbruck, Austria.}
\affiliation{Institute for Theoretical Physics, University of Innsbruck, A-6020 Innsbruck, Austria.}
\email{carlos.gonzalez-ballestero@uibk.ac.at}
\author{Jan Gieseler}
\affiliation{Department of Physics, Harvard University, 17 Oxford Street, Cambridge, MA 02138, USA}
\affiliation{ICFO-Institut de Ciencies Fotoniques, Mediterranean Technology Park, 08860 Castelldefels (Barcelona), Spain}
\author{Oriol Romero-Isart}
\affiliation{Institute for Quantum Optics and Quantum Information of the
Austrian Academy of Sciences, A-6020 Innsbruck, Austria.}
\affiliation{Institute for Theoretical Physics, University of Innsbruck, A-6020 Innsbruck, Austria.}

\begin{abstract}

We theoretically show how to strongly couple the center-of-mass motion of a micromagnet in a harmonic potential to one of its acoustic phononic modes. The coupling is induced by a combination of an oscillating magnetic field gradient and a static homogeneous magnetic field. The former parametrically couples the center-of-mass motion to a magnonic mode while the latter tunes the magnonic mode in resonance with a given acoustic phononic mode. 
The magnetic fields can be adjusted to either cool the center-of-mass motion to the ground state, or to enter into the strong quantum coupling regime.
The center-of-mass can thus be used to probe and manipulate an acoustic mode, thereby opening new possibilities for out-of-equilibrium quantum mesoscopic physics. Our results hold for  experimentally feasible parameters and apply to levitated micromagnets as well as micromagnets deposited on a clamped nanomechanical oscillator.
\end{abstract}

\maketitle

In quantum optomechanics, coupling the mechanical mode of a macroscopic object to a low entropy and narrow mode of the electromagnetic field has enabled ground-state cooling of micromechanical oscillators both in the optical and in the microwave regime
\cite{TeufelNature2011,ChanNature2011}. In addition, strong optomechanical coupling has allowed to generate entanglement between micromechanical and electromagnetic modes \cite{PalomakiScience2013,RiedingerNature2016}, as well as entanglement between remote micromechanical oscillators \cite{RiedingerNature2018}. In this article, we propose a novel analog to linearized quantum optomechanics \cite{AspelmeyerRevModPhys2014} that does not require to couple the mechanical oscillator to an electromagnetic field mode. Instead, we propose to couple the micromechanical oscillator to its internal quantum degrees of freedom. 
 
\begin{figure} 
	\centering
    	\includegraphics[width=\linewidth]{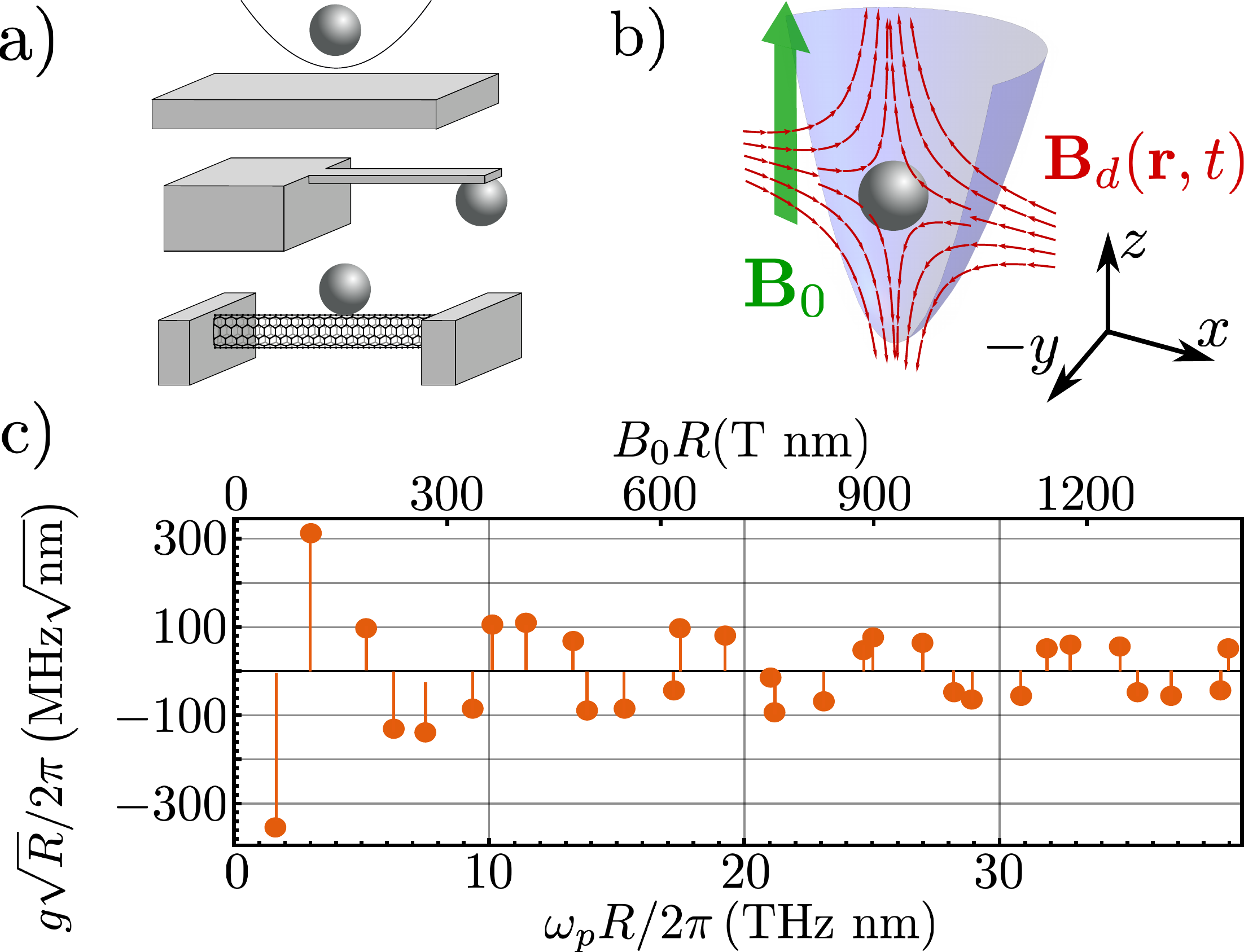}
	\caption{a) \& b) Schematic illustration of our proposal. c) magnetoelastic coupling strength between the Kittel magnon and the first $30$ acoustic spheroidal modes with angular (azimuthal) mode number 2 (1) versus acoustic mode frequency.
	The upper scale indicates the $B_{0}$ needed to tune magnon and phonon in resonance. All axes are normalized to be independent on the micromagnet radius $R$. Parameters correspond to Yttrium-Iron-Garnet (YIG): $\rho_m=5170~\text{kg}/ \text{m}^{3}$, $M_S=5.87\times10^5$ A/m, $|\gamma|=1.76\times10^{11}$T$^{-1}$s$^{-1}$~\cite{ZhangSciAdv2016,StancilBook2009}.}
	\label{figsystem}
\end{figure}

 Our proposal considers the center-of-mass motion of a micromagnet, which is either levitated \cite{PratCampsPRAPP2017,RusconiPRL2017,HuilleryArxiv2019,BarowskiJTP1993,DrugeNJP2014}
 or attached to a high-Q micromechanical oscillator 
 \cite{BurgessScience2013,VinanteNatComm2011,ShamsudhinSmall2016,FischerNJP2019,KolkowitzScience2012}, see \figref{figsystem}a. In the presence of properly tuned magnetic fields, we show how the inherent strong magnetoelastic coupling in the micromagnet can be utilized to achieve an {\em acoustomechanical} coupling between the center-of-mass motion of the micromagnet and one of its internal acoustic phononic modes.
We show how both ground state cooling and strong quantum acoustomechanical coupling can be achieved with experimentally feasible parameters. Our proposal thus establishes a method to probe and control collective quantum excitations of a levitated nanoparticle, thereby opening new possibilities for studying out-of-equilibrium quantum mesoscopic physics, such as, \eg~internal equilibration and radiative cooling with a levitated nanoparticle~\cite{RubioLopezPRB2018}.

We consider a spherical micromagnet of radius $R$ trapped in a harmonic potential, assumed non-magnetic for simplicity. The micromagnet interacts with an external magnetic field,  which has a homogeneous component $\mathbf{B}_{0} = B_{0}\mathbf{e}_z$, and an oscillating gradient $\mathbf{B}_d(\mathbf{r},t) = b_g(-x\mathbf{e}_x+z\mathbf{e}_z)\cos(\omega_d t)$ with $b_gR \ll B_0$ (\figref{figsystem}b). 
The Hamiltonian describing the dynamics of the relevant coupled degrees of freedom of the micromagnet is given  by
\begin{multline}\label{3modeinitialH}
            \frac{\hat{H}(t)}{\hbar} = \omega_{x}\hat{b}^\dagger\hat{b} + \omega_m\hat{s}^\dagger\hat{s} + \omega_p\hat{a}^\dagger\hat{a} + g\left(\hat{s}^\dagger\hat{a} + \hat s \adop \right) 
            \\
             +G_x\cos(\omega_d t)\left(\hat{s}^\dagger+\hat{s}\right)(\hat{b}^\dagger+\hat{b}).
\end{multline}
The first three terms describe, using bosonic operators, the free dynamics of the center-of-mass motion along the $x$-axis ($\bop$), the magnonic mode ($\hat s$), and an acoustic mode ($\aop$) whose frequency is close to the magnonic mode frequency. The fourth term corresponds to the magnetoelastic coupling between the magnon and the acoustic phonon, whereas the last term describes the time-dependent coupling between the magnon and the center-of-mass motion due to the inhomogeneous drive $\mathbf{B}_d(\mathbf{r},t)$. Similar field inhomogeneities have been exploited to couple internal and external degrees of freedom in levitated nanodiamonds \cite{ConanglaNanoLett2018,RahmanNatPhot2017,DelordPRL2018,HoangNatComm2016}.  

The Hamiltonian \eqnref{3modeinitialH} is obtained as follows, see~\cite{GonzalezBallesteroarXiv2019} for further details. First, one quantizes spin waves in a spherical micromagnet around the equilibrium point induced by $\BB_0$, employing the dipolar, isotropic, and magnetostatic approximations \cite{StancilBook2009,WalkerPhysRev1957,Fletcher59,MillsJMMM2006}, which are valid for micromagnet sizes 
$ 10 \text{nm} \lesssim R\lesssim $ 1 $\text{cm}$. We focus on the Kittel magnonic mode, which corresponds to a homogeneous magnetization precessing around the $z-$axis  with frequency  $\omega_m = \vert\gamma\vert B_0$, where $\gamma$ is the gyromagnetic ratio. Second, one quantizes linear  elastic waves in a sphere~\cite{Eringenbook,LambPLMS1881}, obtaining analytical expressions for the acoustic modes with frequencies proportional to $R^{-1}$. Within this linear theory, elastic waves and center-of-mass motion are uncoupled. Third, the magnetoelastic interaction is calculated~\cite{Landaubook,ZhangSciAdv2016}, and, for nano- and micrometer-sized magnets, the leading contribution is of quadratic form. In addition, one obtains selection rules showing that the Kittel mode only couples to acoustic phononic modes of the family $S_{n21}$, that is, spheroidal modes with fixed angular (azimuthal) mode index 2 (1) and arbitrary radial positive integer index $n$. By tuning $B_0$ such that the Kittel magnon frequency $\w_m$ is close to the resonance frequency of an acoustic mode $S_{n21}$, and using the rotating wave approximation, valid for sufficiently small coupling rate and magnon-phonon detuning, the magnetoelastic interaction is described by the beam-splitter form given in \eqnref{3modeinitialH}. The scaled coupling rate ($g \propto R^{-1/2}$) is shown in \figref{figsystem}c for  $n=1,\ldots, 30$, which also evidences the well-discretized spectrum of the acoustic phonons.
Finally, the interaction between the center-of-mass motion and the Kittel magnon, namely the last term in \eqnref{3modeinitialH}, is obtained from the micromagnetic energy density term accounting for the magnetic dipolar coupling with $\BB_d(\rr,t)$. Assuming the motional amplitude of the center of mass to be much smaller than $B_0/b_g$, the $R-$independent coupling rate is given by $G_x = b_g V \mathcal{M}_K x_{0}/(2\hbar)$, where $V$ is the volume of the micromagnet,  $\mathcal{M}_K =\sqrt{\hbar\vert\gamma\vert M_S/2V}$  the zero-point magnetization of the Kittel magnon,  $M_S$ the saturation magnetization \cite{StancilBook2009,MillsJMMM2006}, and $x_{0} = \left[2\rho_m V\omega_{x}/\hbar\right]^{-1/2}$  the zero-point motion of the center-of-mass oscillation along the $x-$axis, where $\rho_m$ is the mass density of the micromagnet~\footnote{Note that, in the case of a micromagnet deposited on a sufficiently massive micromechanical oscillator, the  quantity $\rho_m V$ should be substituted by the effective mass of the oscillator}. 

The system dynamics is described by the master equation $\dot{\rho} = (i\hbar)^{-1}[\hat{H}, \hat \rho] + \mathcal{L}[\hat \rho]$, where $\hat \rho$ is the density operator and $\mathcal{L}[\hat \rho] = \mathcal{L}_m[\hat \rho]+\mathcal{L}_p[\hat \rho]+\mathcal{L}_{x}[\hat \rho]$ accounts for the unavoidable dissipation. Such dissipators, for $j=m,p,x$, are given by $\mathcal{L}_j[\hat \rho] = \gamma_j[(\bar{n}_j+1)L_{\hat{o}_j}+\bar{n}_jL_{\hat{o}^\dagger_j}]$, where for compactness we define $\{\hat{o}_m,\hat{o}_p,\hat{o}_{x}\}\equiv \{\hat{s},\hat{a},\hat{b}\}$ and $L_{\hat{o}}[\hat \rho]\equiv\hat{o}\hat \rho\hat{o}^\dagger-\{\hat{o}^\dagger\hat{o},\hat \rho\}/2$. We have introduced the decay rate $\gamma_j$ and thermal occupation number $\bar{n}_j=(\exp\left[\hbar\omega_j/k_B T_{e,j}\right]-1)^{-1}$, where  $k_B$ is the Boltzmann constant and $T_{e,j}$ the temperature of the thermal environment of each degree of freedom.
The above master equation is quadratic and can thus be solved exactly. It is convenient to define mechanical and acoustic quality factors as $Q_x \equiv \omega_x/\gamma_x$ and $Q_{p} \equiv \omega_{p}/\gamma_{p}$, respectively. 
Experimental values for $Q_x$ exceed $Q_x \gtrsim 10^8$ both in nanofabricated resonators 
\cite{NortePRL2016,ReinhardtPRX0216,GhadimiScience2018,MasonNatPhys2019}
and levitated systems \cite{GieselerNatPhys2013,GieselerPRL2012}. Regarding $Q_p$, unusually high values ($Q_p \approx 10^5-10^7$) have been reported in millimeter-sized Yttrium-Iron-Garnet (YIG) spheres \cite{ZhangSciAdv2016,LeCrawPRL1961}, but no measurements have been performed for isolated micromagnets of the sizes considered in this article. 
However, for sufficiently isolated mechanical microresonators, $Q_p$ is known to be limited by indirect interactions with other acoustic modes, and reaches values up to $Q_p \gtrsim  5\times10^{10}$ \cite{ZhangSciAdv2016,MaccabearXiv2019} when consecutive acoustic modes are far detuned ($\gtrsim $GHz). Therefore, one might expect values of $Q_p$ as high as $\sim 10^{10}$ in our system.

To discuss the center-of-mass dynamics, it is convenient to diagonalize  the magnon-phonon Hamiltonian through a Bogoliubov transformation  $\omega_m\hat{s}^\dagger\hat{s} + \omega_p\hat{a}^\dagger\hat{a} + g\left(\hat{s}^\dagger\hat{a} + \text{H.c.}\right) = \sum_{\alpha=1,2}\omega_\alpha\hat{c}_\alpha^\dagger\hat{c}_\alpha$. The new normal modes are hybrid magnon-phonon excitations given by
$\hat{c}_1=(\hat{s}-\chi\hat{a})/\mathcal{N}$ and $\hat{c}_2=-(\chi\hat{s}+\hat{a})/\mathcal{N}$, 
where $\mathcal{N}\equiv\sqrt{1+\chi^2}$, $\chi \equiv -2g/[\Delta-(\Delta^2+4g^2)^{1/2}]$, and $\Delta \equiv \omega_m-\omega_p$. The phonon(magnon) fraction in mode $\hat c_1$($\hat c_2$) is given by $(\chi/\mathcal{N})^2$. Both the factor $\chi \in [0,\infty)$ and the eigenfrequencies $2\omega_\alpha = \omega_m+\omega_p+(-1)^\alpha(\Delta^2+4g^2)^{1/2}$ are fully tunable through the external field $B_0$. In terms of the normal modes and in the rotating frame $\hat{U}(t) = \exp (i\omega_d t\sum_{\alpha}\hat{c}_\alpha^\dagger\hat{c}_\alpha)$, the Hamiltonian \eqnref{3modeinitialH} reads 
\begin{multline}\label{finalH}
\frac{\hat{H}}{\hbar} = \omega_x\hat{b}^\dagger\hat{b}+ \sum_{\alpha=1,2} \Delta_\alpha\hat{c}^\dagger_\alpha\hat{c}_\alpha \\ + (\hat{b}^\dagger +\hat{b}) \sum_{\alpha=1,2} (G_{x\alpha}\hat{c}^\dagger_\alpha + \hc ),    
\end{multline}
after a rotating wave approximation, valid for $\omega_d \gg \vert G_x\vert/4,\omega_x$. Here $\Delta_\alpha \equiv \omega_\alpha-\omega_d$, and the couplings are renormalized to $G_{x1}=G_x/(2\mathcal{N})$, and $G_{x2} = -\chi G_{x1}$. In terms of the normal modes, the dissipators take the form $\mathcal{L}_p[\hat\rho]+\mathcal{L}_m[\hat\rho] = \mathcal{L}_{12}[\hat\rho] +\sum_{\alpha=1,2} (\gamma_{\alpha+}L_{\hat{c}_\alpha^\dagger}[\hat\rho]+\gamma_{\alpha -}L_{\hat{c}_\alpha}[\hat\rho])$. The $\mathcal{L}_{12}$ term describes an incoherent interaction which can be neglected, under a rotating wave approximation, for micromagnet radii $R\lesssim 10\mu$m \cite{GonzalezBallesteroarXiv2019}.  The corresponding rates in the remaining terms are given by
$\gamma_{1\xi}=\Gamma_{m\xi}+\chi^2\Gamma_{p\xi}$ and $\gamma_{2\xi}=\chi^2\Gamma_{m\xi}+\Gamma_{p\xi}$, with $\Gamma_{m\xi}\equiv\gamma_m(\bar{n}_m+\delta_{\xi -})/\mathcal{N}^2$
and
$\Gamma_{p\xi}\equiv\gamma_p(\bar{n}_p+\delta_{\xi -})/\mathcal{N}^2$. Here $\xi = +,-$ and $\delta_{\xi \xi'}$ is a Kronecker delta. We define the linewidth of the mode $\hat c_1$ ($\hat c_2$) as   $\gamma_1 \equiv (\gamma_m+\gamma_p\chi^2)/\mathcal{N}^2$ ($\gamma_2\equiv (\gamma_m \chi^2+\gamma_p)/\mathcal{N}^2$). Note that in terms of the normal modes, the center of mass is coupled to two independent, largely detuned modes, as $\Delta_1-\Delta_2 = 2(\Delta^2+4g^2)^{1/2} \gg\omega_x$ for typical mechanical frequencies.
To maximize the acoustomechanical interaction, the magnetic field parameters $B_0$ and  $\w_d$ are adjusted such that mode $\hat c_2$ is in resonance with the center of mass motion ($\Delta_2 = \w_x$) and $\chi=10^{-2}$, so that $\hat c_2$ is mainly ($\approx 99.99\%$) acoustic~\footnote{Note that although the roles of $\hat c_1$ and $\hat c_2$ could be exchanged by tuning the former in resonance to the center of mass and choosing a large $\chi$, the present choice is preferred as it requires smaller fields $B_0$ \cite{GonzalezBallesteroarXiv2019}.}. The associated decrease in the coupling between $\hat c_2$ and the center of mass, $G_{x2}\propto\chi$, can be independently compensated by increasing the field gradient $b_g$. In this way, we form a quasi-two-mode acoustomechanical system where the mechanical motion of the micromagnet is coupled to the mainly (99.99\%) acoustic $\hat c_2$ mode, which plays the role of the electromagnetic mode in optomechanics~\cite{AspelmeyerRevModPhys2014}.

\begin{figure} 
	\centering
    	\includegraphics[width=\linewidth]{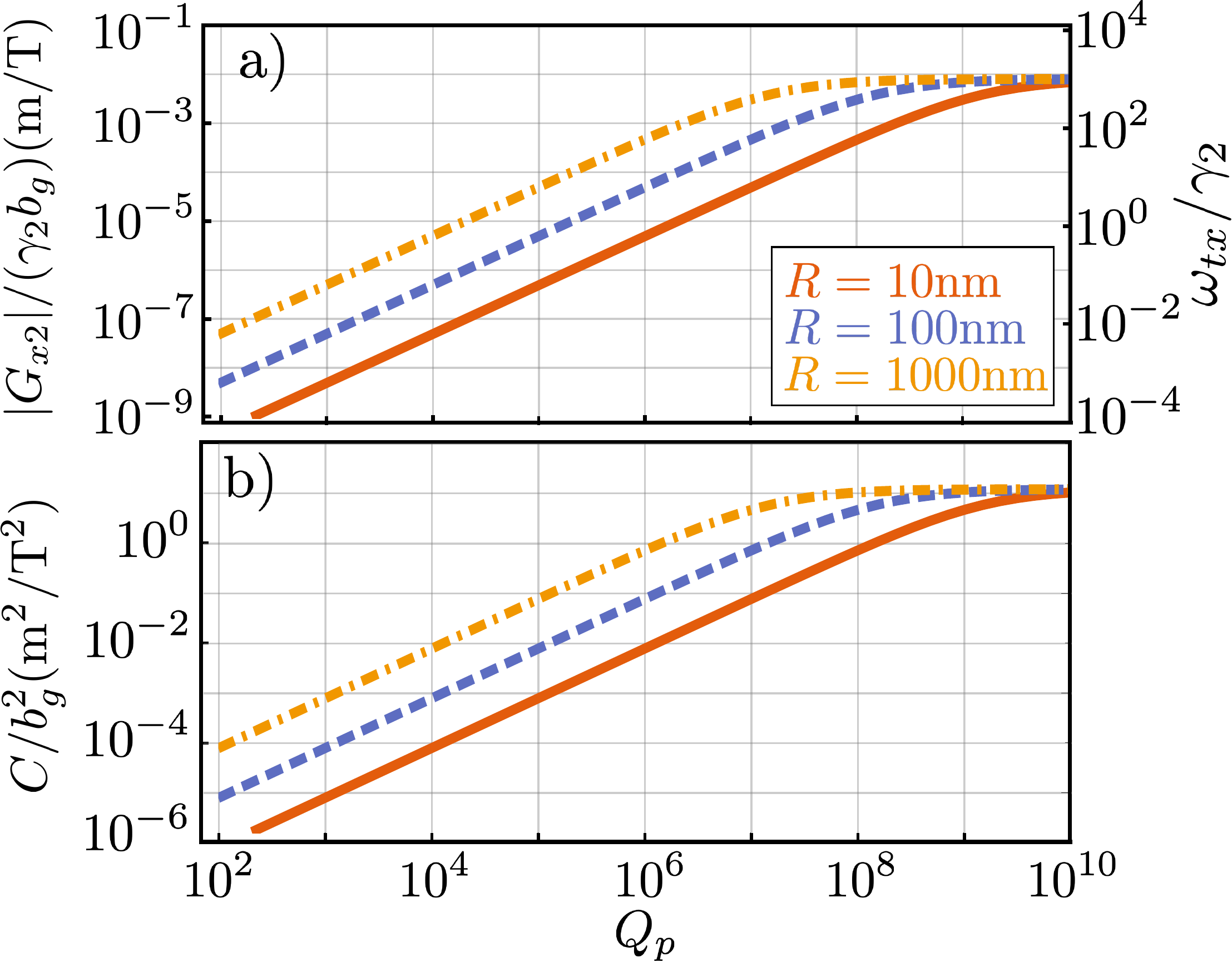}
	\caption{Relevant acoustomechanical parameters for $Q_x = 10^8$ versus acoustic quality factor $Q_p$. The dependence with the field gradient $b_g$ has been factored out explicitly, so the figures are $b_g-$independent. a) Coupling (left) and mechanical frequency (right) normalized to linewidth of mode $\hat{c}_2$. b) Single-phonon cooperativity $C\equiv4G_{x2}^2/\gamma_2\gamma_x$. In both panels we fix $\chi = 10^{-2}$ by applying an external static field $B_{0} \approx\{5.4,0.45,0.018\}$T for $R=\{10,10^2,10^3\}$nm, respectively.}\label{figParams}
\end{figure}

In order to explore the physical regimes that our proposed acoustomechanical system can achieve, we plot the ratio $\vert G_{x2}\vert/\gamma_2 \propto b_g$ in \figref{figParams} (panel a, left axis), the ratio $\omega_x/\gamma_2$ (panel a, right axis), and the cooperativity $C\equiv 4G_{x2}^2/(\gamma_2\gamma_x) \propto b_g^2$ (panel b), as a function of the  acoustic quality factor $Q_p$ and for three values of the micromagnet radius $R$. Hereafter, we consider the lowest-order ($S_{121}$) phonon, material parameters for YIG, and fix $\omega_x = 2\pi\times 200$kHz and $\gamma_m =2\pi\times 1$MHz \cite{TabuchiPRL2014}. All the quantities in \figref{figParams} increase initially as a function of $Q_p$, as the linewidth $\gamma_2 \approx \gamma_m \chi^2 + \omega_p/Q_p$ is reduced, and saturate for large $Q_p$ where $\gamma_2\to \chi^2\gamma_m$ becomes magnon-limited.
The system resides both in the resolved sideband regime ($\w_x > \gamma_2$) and the high cooperativity ($C>1$) regime even at moderate $Q_p \sim 10^6$ and magnetic field gradients $b_g\sim 5$T/m. 
The strong coupling regime ($\vert G_{x2}\vert>\gamma_2$) can also be attained for a wide range of $Q_p$ and feasible gradients $b_g\sim 10^3-10^4$T/m. Moreover, the system can reach the strong quantum cooperativity regime $C/(\bar{n}_x\bar{n}_p) > 1$ allowing for coherent quantum state transfer between mechanical motion and acoustic phonons~\cite{AspelmeyerRevModPhys2014}. Indeed, at cryogenic temperatures ($T_{e,j}=100$mK), the product $\bar{n}_x\bar{n}_p<10^4$ and $C/(\bar{n}_x\bar{n}_p) >1 $ can be achieved for $b_g\gtrsim10^3$T/m and $Q_p\gtrsim 10^6$ for all radii in Fig. \ref{figParams}. At room temperature, attaining such regime is more challenging and only feasible for small $R$ at gradients $\gtrsim 10^4$T/m. 
\figref{figParams} highlights that our acoustomechanical system can be tuned into the resolved-sideband, the high-cooperativity, and either the weak or the strong coupling regime with experimentally accessible parameters. 
This versatility enables a range of applications, which we will discuss in the following.

\begin{figure} 
	\centering
    	\includegraphics[width=\linewidth]{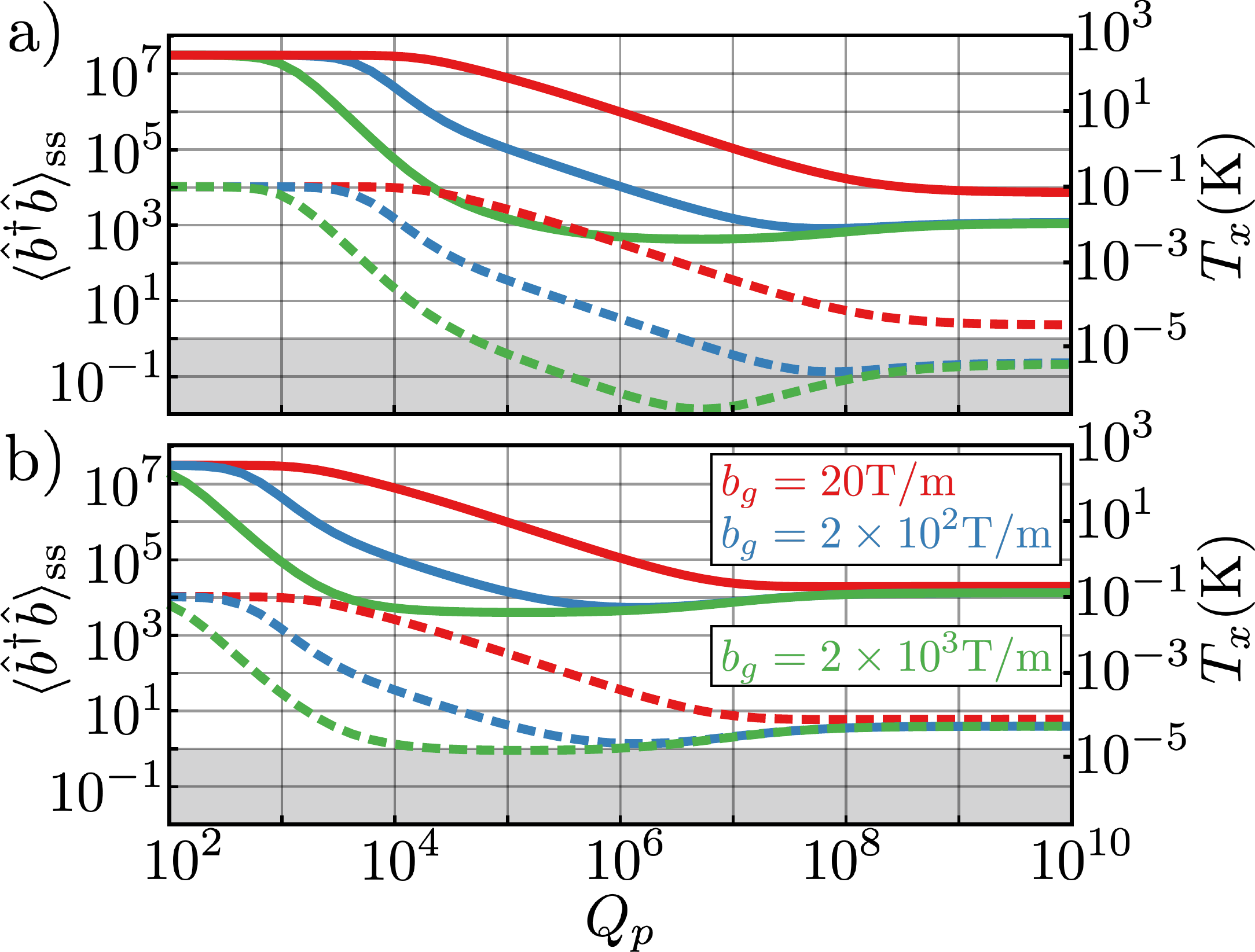}
	\caption{Steady-state center of mass occupation ($Q_x=10^8$) versus acoustic quality factor, for $R=100$nm (a) and $R=1\mu$m (b) and three values of the magnetic gradient $b_g$. Solid and dashed lines indicate results at room ($T_{e,j}=300$K) and cryogenic temperatures ($T_{e,j}=100$mK) respectively. The shaded area indicates the ground state cooling region  $\langle\hat{b}^\dagger\hat{b}\rangle_{\rm ss} <1$. The right axes indicate the steady-state center of mass temperature $k_BT_x\approx \hbar\omega_x\langle\hat{b}^\dagger\hat{b}\rangle_{\rm ss}$ ($\omega_x = 2\pi\times200$kHz).}\label{figCooling}
\end{figure}

Efficient center-of-mass cooling can be achieved in the resolved-sideband, high-cooperativity, and weak-coupling regime~\cite{GenesPRA2008,MarquardtPRL2007,WilsonRaePRL2007}.
By solving the quadratic master equation exactly, the  steady-state occupation of the center of mass, $\langle \hat{b}^\dagger\hat{b}\rangle_{\rm ss}$, can be evaluated. \figref{figCooling}a (\figref{figCooling}b) shows $\langle \hat{b}^\dagger\hat{b}\rangle_{\rm ss}$ for $R=100$nm ($R=1\mu$m) at $Q_x=10^8$ and different field gradients $b_g$, for both a room temperature environment ($T_{e,j}=300$K, solid lines) and cryogenic conditions ($T_{e,j}=100$mK, dashed lines). For small acoustic quality factors, the cooling is inefficient as $C<1$ (see \figref{figParams}). When $Q_p$ increases above a certain value (which depends on $R$ and $b_g$, see \figref{figParams}) the $C>1$ regime is reached and center-of-mass cooling is observed. Notice that ground-state cooling, $\langle\hat{b}^\dagger\hat{b}\rangle_{\rm ss} <1$, is achieved for both micromagnet sizes at  $T_{e,j}=100$mK. For sufficiently high $Q_p$, however, the cooling becomes less efficient as the system enters the strong coupling regime ~\cite{GenesPRA2008,MarquardtPRL2007,WilsonRaePRL2007}. The lowest occupations in both panels of \figref{figCooling}, \ie the minima of the green dashed lines, correspond to parameters at which the mechanical sidebands are very well resolved ($\omega_x/\gamma_2\sim 10^2$, see \figref{figParams}), and thus  cooling is limited by other factors.
For $R=100$nm, the lowest occupation $\langle\hat{b}^\dagger\hat{b}\rangle_{\rm ss,min} \approx 0.014$ at $b_g=2\times10^3$T/m is cooperativity-limited. It can, thus, be reduced by increasing either $b_g$ or $Q_x$. 
In contrast for $R=1\mu$m, $\langle\hat{b}^\dagger\hat{b}\rangle_{\rm ss,min} \approx 0.89$ at $b_g=2\times10^3$T/m is limited by the entropy of the acoustic phonon bath through the phonon occupation $\bar{n}_p$~\footnote{The linear entropy of a thermal state is given by $S_L=\text{Tr}[1-\hat{\rho}^2]=\bar{n}_p/(\bar{n}_p+1/2) \approx 2\bar{n}_p$ for low $\bar{n}_p$.}, which decreases with $R$ due to the reduction of the acoustic frequency. Note that further cooling is still possible in this case by tuning the Kittel mode  close to resonance with a higher order (lower entropy) acoustic phonon \cite{GonzalezBallesteroarXiv2019}, see \figref{figsystem}c. As evidenced by these results, coupling the micromagnet motion to its built-in internal resonators (i.e. phonons) allows for cavity-less cooling of the motion, a specially promising prospect for levitated micromagnets for which neither optical cooling nor microwave optomechanical cooling are efficient, due to absorption and weak coupling rates respectively.
 
 \begin{figure} 
	\centering
    	\includegraphics[width=\linewidth]{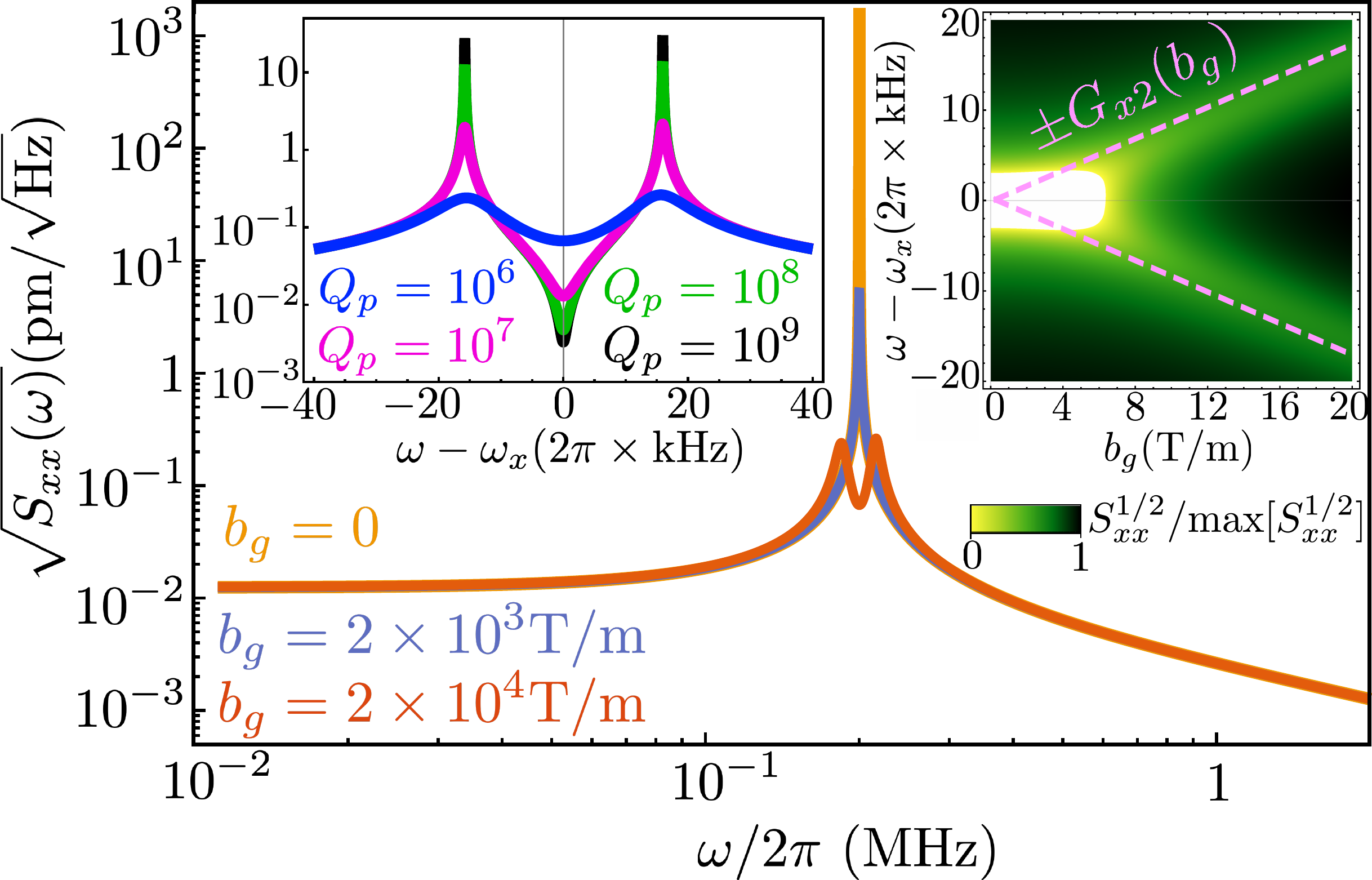}
	\caption{Power spectral density $S_{xx}(\omega)$, for $Q_p=10^6$, $Q_x=10^5$, $R=100$nm, and $T_{e,j}=300$K, at three different values of the field gradient $b_g$. Left inset: peak splitting at $b_g=10^4$T/m, for different acoustic quality factors $Q_p$. Right inset: normalized power spectral density as a function of $b_g$ and detuning $\omega - \omega_x$. The dashed lines indicate the function $G_{x2}(b_g)$. Strong coupling is reached at $b_g\approx10^4$T/m.
	}\label{figPSD}
\end{figure}
 
The strong and tunable acoustomechanical interaction also allows to probe the acoustic phonons by measuring the mechanical displacement of the center of mass.
\figref{figPSD} shows the power spectral density of the the center-of-mass motion $S_{xx}(\omega) \equiv (2\pi)^{-1}\int_{-\infty}^{\infty}d\tau\langle\hat{x}(0)\hat{x}(\tau)\rangle_{\rm ss} e^{i\omega\tau}$,
where $\hat{x} \equiv x_0(\hat{b}^\dagger + \hat{b})$, for $R=100$nm and moderate quality factors $Q_x=10^5$ and $Q_p=10^6$. For these parameters, 
$\omega_x /\gamma_2 \approx 10$ and $2\vert G_{x2}\vert/\gamma_2\approx 10^{-4}b_g$ (see \figref{figParams}a).
We distinguish the three possible regimes in \figref{figPSD}, namely the zero-coupling regime ($b=0$), where the single peak at $\omega=\omega_x$ and width $ \gamma_x \approx 2\pi\times 2$Hz indicates a freely evolving center-of-mass motion; the weak coupling regime ($b_g=2\times10^{3}$T/m), characterized by a reduced peak, \ie by cooling of the mechanical motion; and the strong coupling regime ($b_g= 2\times10^4$T/m), where the peak splits into two~\cite{GroblacherNature2009}. The latter splitting is induced exclusively by the mode $\hat c_2$, \ie by acoustic phonons, as evidenced by the two insets of \figref{figPSD}. In the left inset we observe how the signal increases with the acoustic quality factor, up to the magnon-limited saturation point ($Q_p\sim 10^9$) (see also \figref{figParams}). The right inset shows that the mode splitting is well approximated by the function $2\vert G_{x2}(b_g)\vert$. The strong coupling crossover $2\vert G_{x2}(b_g)\vert/\gamma_2=1$  is at $b_g\approx 10^4$T/m for the chosen parameters.
Measuring the peak splitting due to center-of-mass hybridizing with the acoustic mode requires to resolve the thermal motion of the center-of-mass mode. According to the results in \figref{figPSD}, this lies well within the sensitivity range of most state-of-the-art ultra-sensitive micromechanical sensors~\cite{RossiNanoLett2019,BraakmanNanotech2019,JainPRL2016,GieselerNatPhys2013,BrawleyNatCom2016, WeberNatCom2016}, which can even resolve motion on the quantum level~\cite{MasonNatPhys2019,WollmanScience2015,LecocqPRX2015}. Thus, the acoustic-induced mode splitting is experimentally measurable. 
Let us finally emphasize that, in all the above results, the internal temperature increase of the micromagnet remains low in spite of the magnetic driving, as such driving is largely detuned with respect to the magnon frequency~\cite{GonzalezBallesteroarXiv2019}.

In conclusion, we have shown that the center-of-mass of a micromagnet in a harmonic potential can be coupled, in a strong and tunable way, to one of its internal acoustic phononic modes.
The coupling mechanism can be controlled by external magnetic fields and both ground-state cooling and the strong quantum coupling regime can be achieved. Such a quantum acoustomechanical system opens many possibilities for further research: (i)~exploring the strong quantum cooperativity regime to use an internal acoustic phonon as a quantum memory~\cite{MaccabearXiv2019}, (ii)~preparing the micromagnet in a state where different acoustic modes have different temperatures and probe how they equilibrate~\cite{RubioLopezPRB2018}, (iii)~exploring the regimes where the potentially strong non-linear magnetoelastic interactions~\cite{GonzalezBallesteroarXiv2019} might become relevant, to generate a non-linear hybrid magnon-phonon mode. This mode can act as a qubit and can thus be used to prepare the center-of-mass in a non-Gaussian quantum state. Last but not least, in the context of levitated nanoparticles, our work highlights the important fact that nanoparticles are not point objects with only external degrees of freedom, but complex particles with internal degrees of freedom that can be harnessed in the quantum regime.

C.~G.~B.  and J.~G. acknowledge support from the European Union (PWAQUTEC, H2020-MSCA-IF-2017, no.~796725 and SEQOO, H2020-MSCA-IF-2014, no.~655369 respectively).


 
\bibliographystyle{apsrev4-1}
\bibliography{bibliography}

\end{document}